\documentclass{icrc2009}

\usepackage{graphicx}   
\usepackage[font=footnotesize]{subfig} 
\usepackage{fixltx2e}
\usepackage{url}
\def\schi{{\sc Hi}\ }
\def\fvw{FVW~190.2+1.1\ }

\newcommand{\shorttitle}[1]%
{\markboth{Proceedings of the 31\MakeLowercase{$^{st}$} ICRC, {\L}\'{o}d\'{z} 2009}{#1} }
\newcommand{\etal}{\MakeLowercase{\textit{et al. }}} 

\begin{document}
\title{VERITAS Observations of a "Forbidden Velocity Wing"}

\author{\IEEEauthorblockN{Jamie Holder\IEEEauthorrefmark{1} 
    for the VERITAS Collaboration\IEEEauthorrefmark{2}}
                            \\
\IEEEauthorblockA{\IEEEauthorrefmark{1}Department of Physics and Astronomy and the Bartol Research Institute, University of Delaware, USA.}  
\IEEEauthorblockA{\IEEEauthorrefmark{2}see R.A. Ong \etal (these proceedings) or \url{http://veritas.sao.arizona.edu/conferences/authors?icrc2009}}}

\shorttitle{J. Holder \etal VERITAS Observations of an FVW}
\maketitle

\begin{abstract}

The H.E.S.S. extended Galactic plane survey \cite{chavez09} revealed
the presence of a new extended TeV gamma-ray source, HESSJ1503-582,
with no obvious counterpart at other wavelengths. The source is,
however, coincident with an \schi structure with a velocity
significantly different from that of galactic rotation - a so-called
"Forbidden Velocity Wing" \cite{renaud09}. These structures have been
suggested as the fast moving shells and filaments associated with the
oldest supernova remnants in our galaxy. The detection of TeV
gamma-ray emission from these structures might indicate that supernova
remnants remain efficient particle accelerators for much longer than
is commonly believed. Here we report on recent VERITAS observations of
one of these structures, FVW 190.2+1.1, which shows a clear shell-like
morphology in the \schi maps.
  \end{abstract}


\section{Introduction}

At the Heidelberg Gamma 2008 Symposium, the H.E.S.S. collaboration
reported the results of their extended galactic plane survey,
including the detection of a new extended source ($\sim45$' diameter)
with an integrated flux above 1~TeV of
$6\times10^{-12}$erg~cm$^{-2}$s$^{-1}$, well-fit by a smooth power law
spectrum with photon index
$\Gamma=2.4\pm0.4_{stat}\pm0.2_{sys}$\cite{chavez09, renaud09}. The
source, labelled HESS~J1503-582, has no clear counterpart at other
wavelengths; however, Renaud \etal note the interesting coincidence
with an \schi structure; a so-called ``Forbidden Velocity Wing''. This
motivates an investigation of similar structures with VERITAS in the
Northern hemisphere.

\begin{figure}[!t]
  \centering
  \includegraphics[width=2.5in]{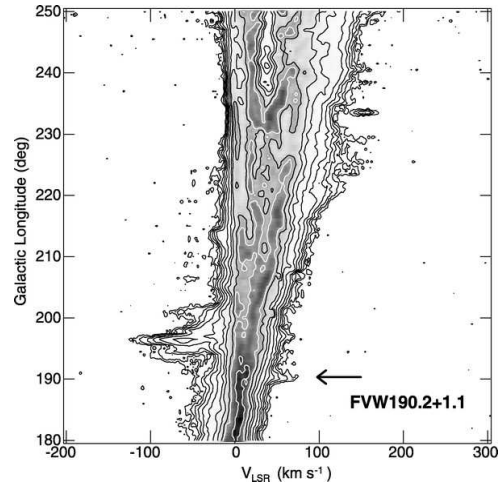}
  \caption{Large-scale Galactic longitude-velocity map of \schi emission at a Galactic latitude of +1.0$^{\circ}$. The location of the forbidden velocity wing FVW~190.2+1.1 is marked. Figure taken from \cite{koo06}.}
  \label{FVW}
\end{figure}

The term ``Forbidden Velocity Wing'' (hereafter FVW) is given to
structures identified through their anomalous velocities, derived from
doppler shifting of \schi line emission. In a plot of velocity ($v$)
against galactic longitude ($l$), these structures appear as spatially
limited ($\leq2^{\circ}$) ``wings'' extending from the bulk of the
galactic \schi emission (Figure~\ref{FVW}), with positive or negative
velocity extensions greater than $\sim20$~km~s$^{-1}$. FVWs are
believed to be sites where kinetic energy has been injected into the
interstellar medium through some violent event. Kang \& Koo
\cite{kang07} have compiled a catalog of 87 FVWs using the
Leiden/Dwingeloo \schi survey data \cite{hartmann97} and the \schi
Southern Galactic Plane Survey data \cite{mcclure01}, which they rank
according to the clarity of the structures in the $l-v$ and $b-v$
velocity diagrams. 33 of the 87 are given the highest rank, one of
which, FVW~319.8+0.3 is co-located with HESS~J1503-582.

\begin{figure*}[th]
  \centering
  \includegraphics[width=5in]{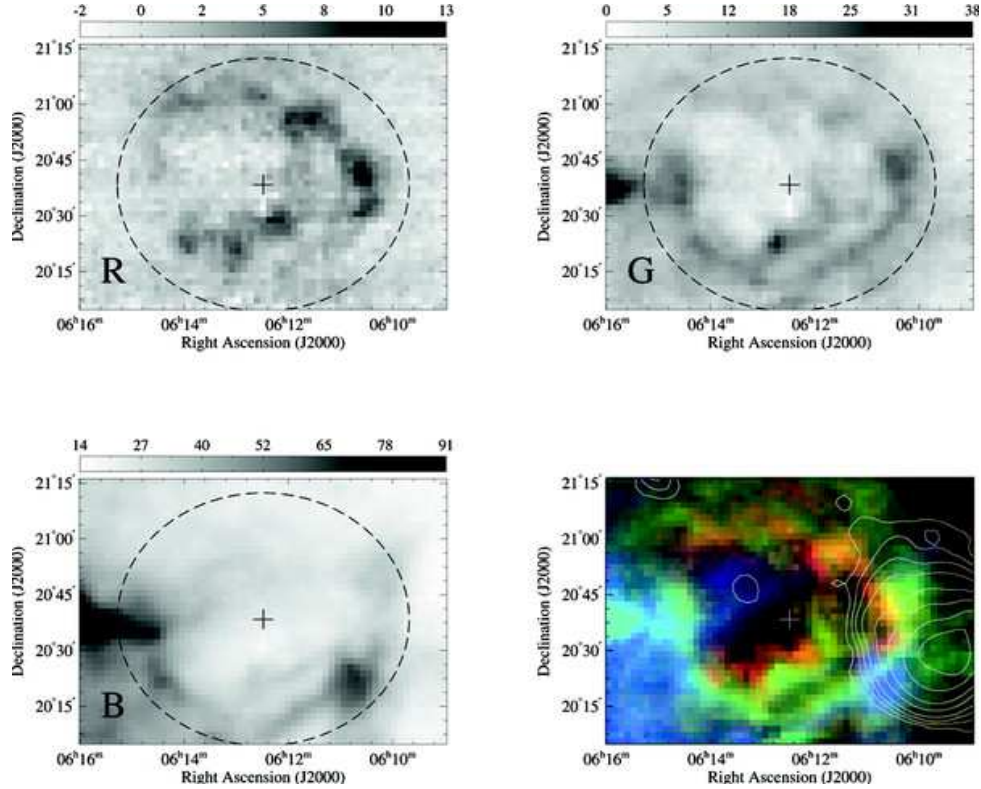}
  \caption{Arecibo \schi images of \fvw, integrated over velocities
    59-79~km~s$^{-1}$ (R), 41-59~km~s$^{-1}$ (G) and 31-41~km~s$^{-1}$
    (B). The lower right panel is a three-colour image generated from
    these. Figure taken from Koo \etal \cite{koo06} - see this reference
    for details.} 
  \label{arecibo}
\end{figure*}

Kang \& Koo discuss a number of potential progenitors for the FVW
structures, among them stellar winds and supernova remnants (SNRs),
both of which are potential TeV gamma-ray sources. In its final
stages, an SNR consists of a rapidly expanding \schi shell, which
might persist after the remnant is too faint to be observed at radio
wavelengths. The number of detected SNR shells in the galaxy is much
lower than the number expected from the supernova rate, presumably
because of these observational limitations. The \schi emission from
another of the highest ranked FVW objects, FVW~190.2+1.1, has been mapped by
Koo \etal with high resolution (3.4') using the 305~m Arecibo
telescope \cite{koo06}. Their results reveal a rapidly expanding
($\sim80$~km~s$^{-1}$) shell structure, invisible at other
wavelengths, which they suggest is consistent with being the shell of
an SNR with an age of $\sim3\times10^5$~years
(Figure~\ref{arecibo}). In these proceedings we report on a search for
counterpart TeV emission from FVW~190.2+1.1 using the VERITAS array.

\section{Observations}
VERITAS \cite{weekes02, holder09} is an array of four imaging
atmosperic Cherenkov telescopes located at the basecamp of the Fred
Lawrence Whipple Observatory near Tucson, Arizona. The array has been
fully operational since mid-2007 and has sensitivity sufficient to
detect a point-like source with 1\% of the steady Crab Nebula flux in
$<50$~hours. Observations cover the energy range from 100~GeV to
$>30$~TeV with an energy resolution of 15-20\% above 300~GeV, and an
angular resolution per gamma-ray photon of $0.1^{\circ}$ at 1~TeV.

Observations of \fvw were made on clear nights between October 26th
and December 28th 2008, at a mean source elevation of
$68^{\circ}$. The \schi source is extended, with an \schi
velocity-dependent diameter of $\sim1^{\circ}$. The centre of the
shell was offset from the centre of the field-of-view by $0.7^{\circ}$
(80\% of the observations) or $0.5^{\circ}$ (20\% of the observations)
to allow simultaneous background estimation (\textit{wobble} mode
\cite{fomin94}). The physical camera field-of-view of VERITAS is
$3.5^{\circ}$ in diameter; \textit{wobble}-mode observations allow us
to map a somewhat larger field-of-view with long
exposures. Observations taken under poor weather conditions were
rejected from the analysis. The final exposure consists of 18.4 hours
of good quality data, with all four telescopes operational.

\section{Results}

The data were analysed using standard VERITAS analysis tools
\cite{acciari_LSIa}. Cherenkov images were first calibrated and
cleaned, then parameterized according to their first and second
moments \cite{hillas85}. The air shower impact position and arrival
direction were then reconstructed using all events wherein at least 3
telescopes recorded a Cherenkov image. The
\textit{mean-reduced-scaled width} and \textit{mean-reduced-scaled
length} parameters were calculated (e.g. \cite{krawczynski06}) and
used for gamma-hadron separation. A further cut was applied on the
arrival direction of the incoming gamma-ray relative to the test
position on the sky ($\theta$). Since the angular extent and spectrum
of the putative TeV source is not well defined, four different sets of
gamma-ray selection cuts were applied, to provide best sensitivity for
both point-like ($\theta<0.122^{\circ}$) and moderately extended
($\theta<0.235^{\circ}$) sources, and for sources with a Crab-like
spectrum (\textit{standard} cuts) and for significantly harder spectra
(\textit{hard} cuts). \textit{Standard} and \textit{hard} cuts differ
only in the minimum number of photo-electrons required ($\sim75$ and
$\sim225$ respectively) for a Cherenkov image to be used in the event
reconstruction. This approach is similar to that typically used in the
analysis of survey data for imaging atmospheric Cherenkov telescopes
\cite{aharonian06, weinstein09}. The background in each test source
region was estimated using the ``ring-background'' method
\cite{berge07}. All results have been verified using two independent
analysis chains.

  \begin{table}[!h]
  \caption{99\% Confidence Upper Limits}
  \label{ULs}
  \centering
  \begin{tabular}{|c|c|c|}
  \hline
   Analysis Cuts &  Upper Limit (F$>$500~GeV)\\ 
                 &   (ph~cm$^-2$~s$^-1$)\\
   \hline 
      Standard spectrum, point source     & 3.6$\times10^{-13}$ \\
      Standard spectrum, extended source  & 4.3$\times10^{-13}$ \\
      Hard spectrum, point source         & 3.9$\times10^{-13}$ \\
      Hard spectrum, extended source      & 3.5$\times10^{-13}$ \\
  \hline
  \end{tabular}
  \end{table}

\begin{figure*}[th]
  \centering
  \includegraphics[width=2.9in]{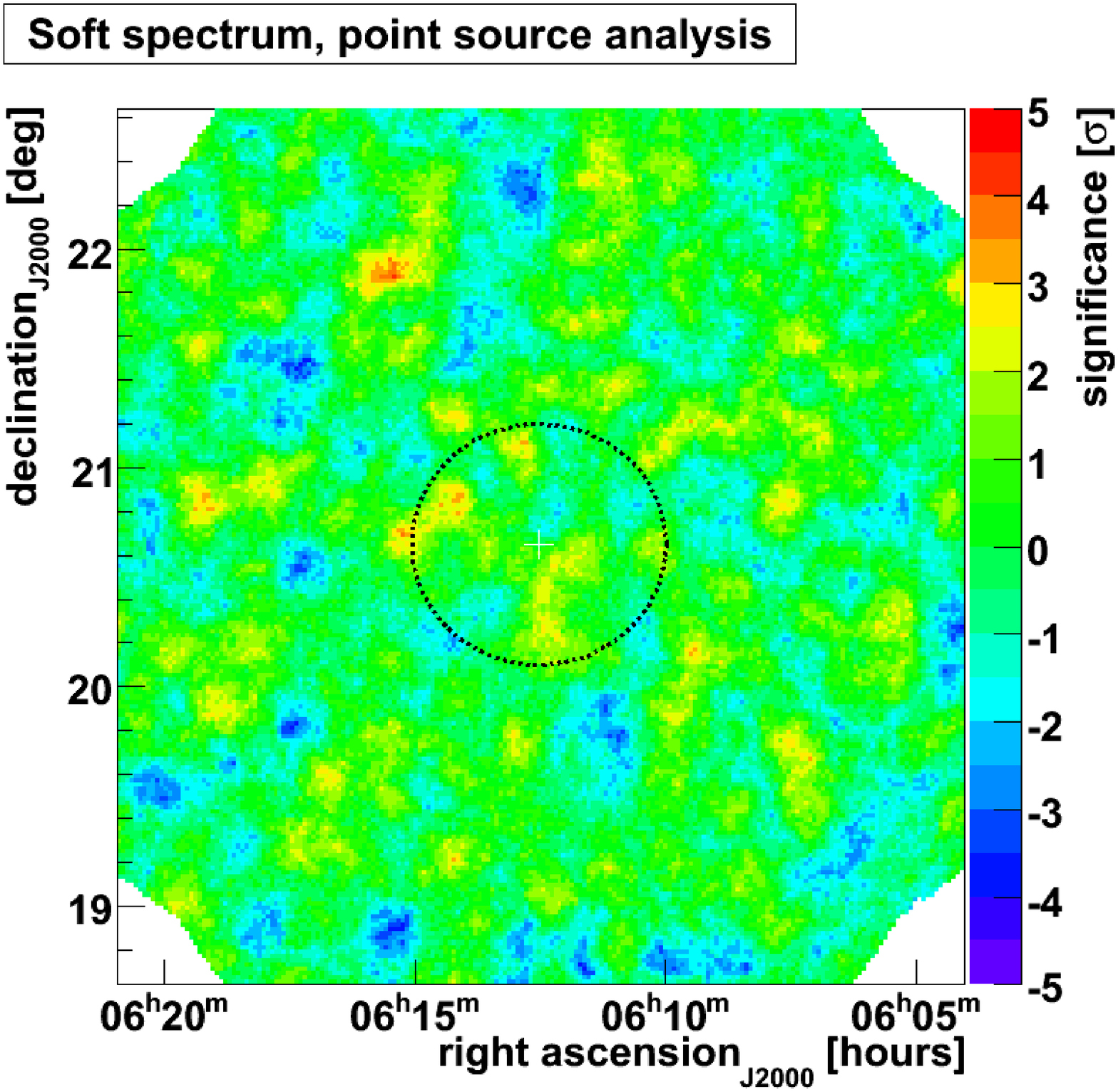}\hspace{0.5cm}\includegraphics[width=2.9in]{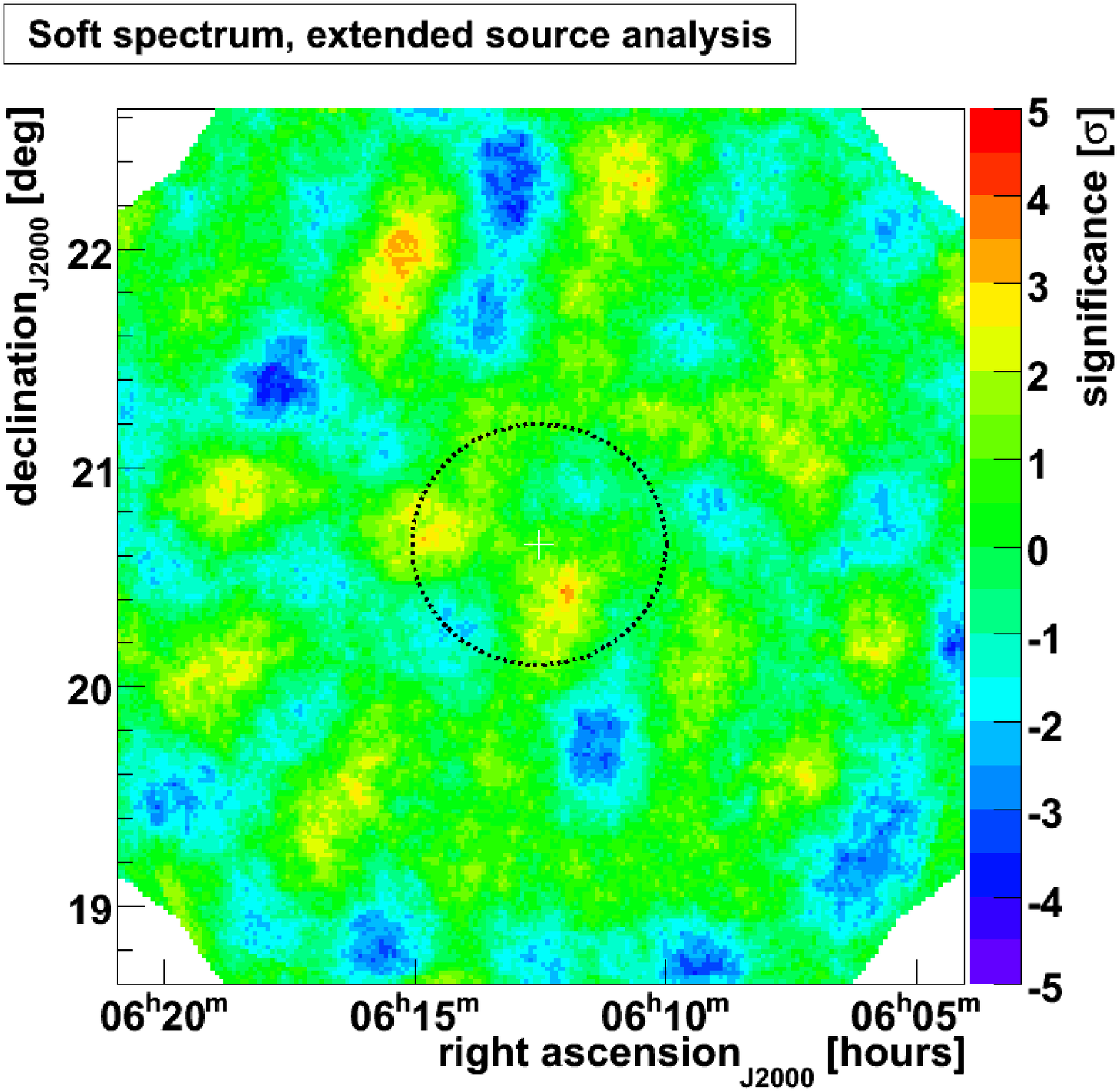}
  \vspace{0.5cm}
  \includegraphics[width=2.9in]{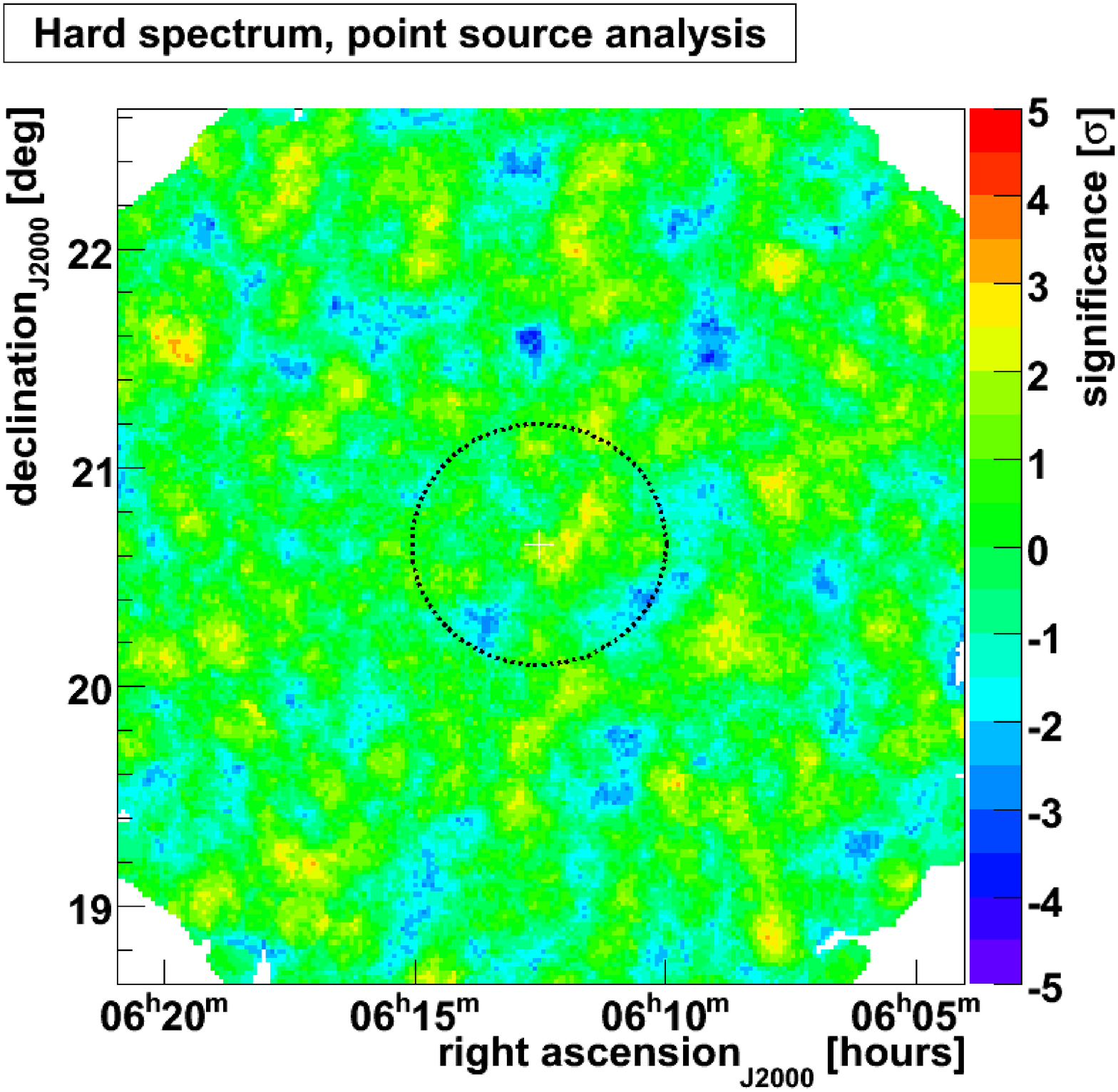}\hspace{0.5cm}\includegraphics[width=2.9in]{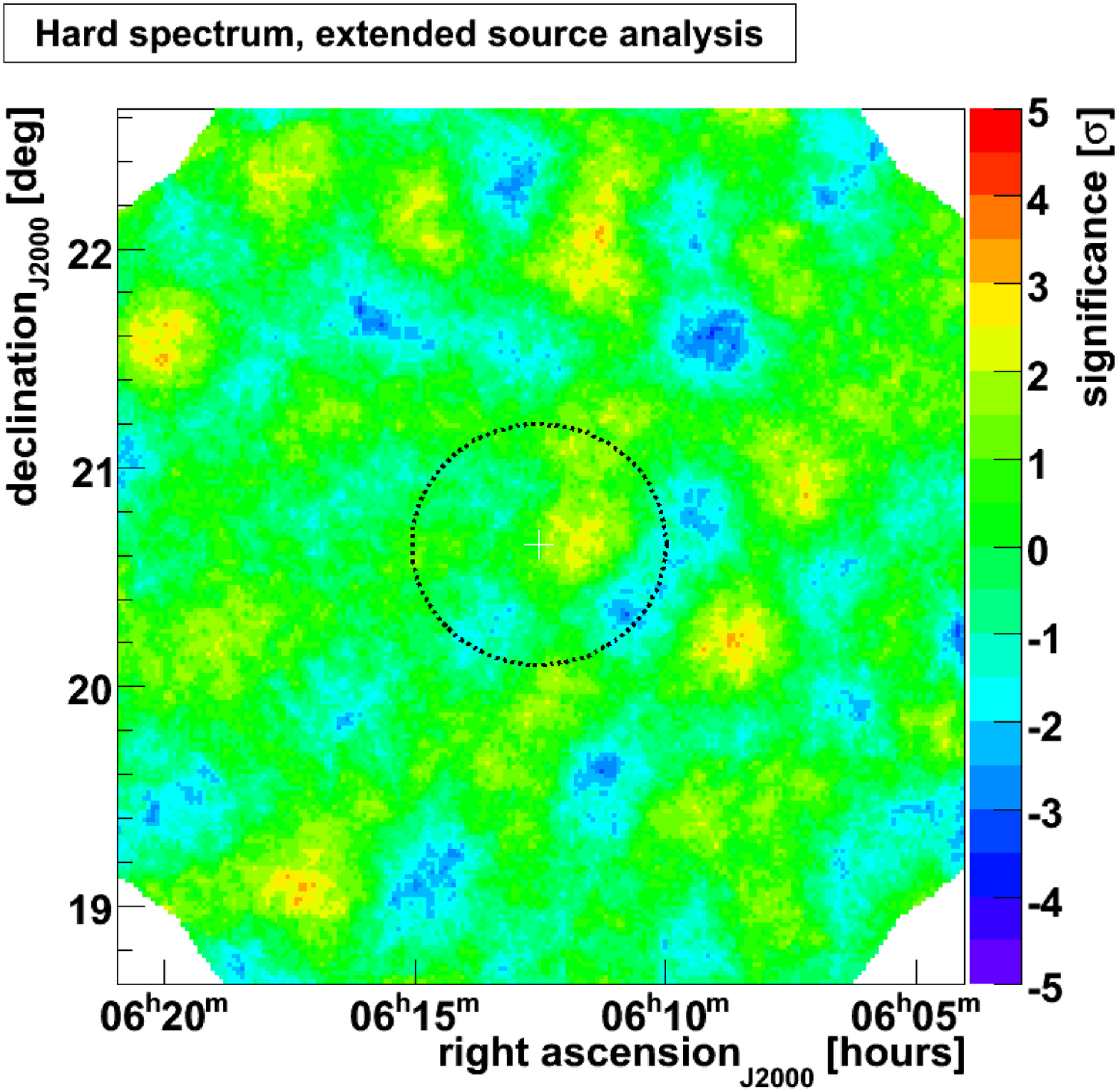}
  \caption{Sky maps of the excess significance for each test position
    in the field around FVW~190.2+1.1. The dashed ellipse indicates the derived
    angular extent of the expanding shell in Koo \etal Maps for each of
    the four different gamma-ray selection cuts are shown as labelled. The results
    are consistent with the expectation for background fluctuations in
    the absence of a gamma-ray source.}
  \label{maps}
\end{figure*}

The resulting sky maps are shown in Figure~\ref{maps}. No significant
excess is observed at any position for any of the four analyses; all
positive excesses are consistent with the expectation for fluctuations
in the absence of any gamma ray source. Table~\ref{ULs} shows the
derived upper limits (calculated using the method of Helene
\cite{helene83}) at the centre of the field-of-view for each cut.

\section{Discussion}

FVW structures are clearly among the more speculative candidates as
counterparts to the unidentified TeV sources. Assuming its
identification as an SNR is correct, Koo \etal estimate a distance to
\fvw of 8~kpc and a dynamical age of $t=3.4\times10^5$~years. TeV
emission from an isolated old remnant such as this is not expected
under conventional models, although Yamazaki \etal \cite{yamazaki06}
predict that the TeV/X-ray flux ratio might increase significantly
as remnants age. Old remnants have already been detected at TeV
energies, for example in the case of W28 \cite{aharonian08}, with an
estimated age of between $0.4 - 1.5\times10^5$~years. In this case the
emission is likely enhanced by interaction with surrounding molecular
clouds. Another, originally unidentified, H.E.S.S. source, HESS
J1713-347 \cite{aharonian08b} is now believed to be associated with a
newly identified shell-type SNR, G353.6-0.7, with an estimated age of
$0.27\times10^5$~years at a distance of $3.2\pm0.8$~kpc \cite{tian08}. 

In summary, we have observed a Forbidden Velocity Wing \schi
structure, FVW~190.2+1.1, that is most likely associated with a supernova
remnant. We see no evidence for TeV gamma-ray emission from this
region, and set upper flux limits at the level of $<1\%$ of the steady
Crab Nebula flux.

\section{Acknowledgements}

This research is supported by grants from the US Department of Energy,
the US National Science Foundation, and the Smithsonian Institution,
by NSERC in Canada, by Science Foundation Ireland, and by STFC in the
UK. We acknowledge the excellent work of the technical support staff
at the FLWO and the collaborating institutions in the construction and
operation of the instrument.

\end{document}